\documentclass[12pt]{article}
\usepackage{epsfig}
\usepackage{amsmath}
\usepackage{amsfonts}
\usepackage{amssymb}

\makeatletter

\renewcommand\maketitle{\par
  \begingroup
    \if@twocolumn
      \ifnum \col@number=\@ne
        \@maketitle
      \else
        \twocolumn[\@maketitle]%
      \fi
    \else
      \newpage
      \global\@topnum\z@   
      \@maketitle
    \fi
    \thispagestyle{plain}\@thanks
  \endgroup
  \setcounter{footnote}{0}%
  \global\let\thanks\relax
  \global\let\maketitle\relax
  \global\let\@maketitle\relax
  \global\let\@thanks\@empty
  \global\let\@author\@empty
  \global\let\@date\@empty
  \global\let\@title\@empty
  \global\let\title\relax
  \global\let\author\relax
  \global\let\date\relax
  \global\let\and\relax
}

\makeatother

\newcommand{\bra}[1]{\ensuremath{\langle{#1}|}}
\newcommand{\ket}[1]{\ensuremath{|{#1}\rangle}}

\renewcommand\[{[\,}

\newcommand{\bea}{\begin{eqnarray}}
\newcommand{\eea}{\end{eqnarray}}
\newcommand{\be}{\begin{equation}}
\newcommand{\ee}{\end{equation}}
\newcommand{\bi}{\begin{itemize}}
\newcommand{\ei}{\end{itemize}}
\newcommand{\bc}{\begin{center}}
\newcommand{\ec}{\end{center}}
\newcommand{\ben}{\begin{enumerate}}
\newcommand{\een}{\end{enumerate}}

\newcommand{\nn}{\nonumber}
\newcommand{\ld}{\ldots}
\newcommand{\cd}{\cdot}
\newcommand{\fr}{\frac{1}{2}}
\newcommand{\fri}{\frac{i}{2}}

\newcommand{\V}{{\cal V}}

\renewcommand{\H}{{\cal H}}

\newcommand{\D}{{\cal D}}

\newcommand{\G}{{\cal G}}

\newcommand{\N}{{\cal N}}
\newcommand{\pa}{\partial}

\renewcommand{\a}{\alpha}
\renewcommand{\b}{\beta}
\renewcommand{\d}{\delta}

\newcommand{\g}{\gamma}
\newcommand{\ep}{\epsilon}

\renewcommand{\l}{\lambda}

\renewcommand{\o}{\omega}
\renewcommand{\r}{\rho}

\renewcommand{\c}{\chi}
\newcommand{\x}{\xi}
\renewcommand{\th}{\theta}

\newcommand{\De}{\Delta}
\newcommand{\Ga}{\Gamma}

\newcommand{\Om}{\Omega}

\let\ssection=\section
\renewcommand{\section}{\setcounter{equation}{0}\ssection}

\begin{document}

\begin{titlepage}

\title{\LARGE Phase Space Representations and Perturbation Theory for
Continuous-time Histories}

\author{
   Aidan Burch\footnote{Current Address: Dept. of Physics, University of Auckland,
   Private Bag 92019, Auckland, New Zealand.}\footnote{Email: aidan.burch@imperial.
   ac.uk} \\[0.5cm]
   {\normalsize The Blackett Laboratory, Imperial College, London SW7 2BZ, UK}}

\date{\today}

\maketitle

\begin{abstract}\normalsize
We consider two technical developments of the formalism of
continuous-time histories. First, we provide an explicit description
of histories of the simple harmonic oscillator on the classical
histories phase space, comparing and contrasting the Q, P and Wigner
representations; we conclude that a representation based on coherent
states is the most appropriate. Second, we demonstrate a generic
method for implementing a perturbative approach for interacting
theories in the histories formalism, using the quartic anharmonic
oscillator. We make use of the identification of the closed-time
path (CTP) generating functional with the decoherence functional to
develop a perturbative expansion for the latter up to second order
in the coupling constant. We consider both configuration space and
phase space histories.

\end{abstract}

\thispagestyle{empty}

\end{titlepage}

\newpage

\section{Introduction}\label{sec:introduction}

The consistent histories approach to quantum mechanics is a
framework for the description of individual (closed) quantum systems
(see \cite{HIS1},\cite{HPOa},\cite{Sav98},\cite{Sav01a}). It
provides a reformulation of quantum physics based on histories,
namely temporally extended propositions for a physical system. One
therefore asks questions about histories of momentum, position,
energy and other variables. The probability information for the
histories is contained within the decoherence functional, which is a
complex valued functional of pairs of histories. In the usual
interpretation, its diagonal elements define probabilities within a
set of histories (usually coarse-grained), provided a specific
consistency condition is satisfied. The basic mathematical objects
of histories theory are therefore different from those of the
standard formulation, even though in the cases of interest the
former can be constructed from the latter.

This paper deals with two specific technical issues of the history
formalism that have not been fully developed in the relevant
bibliography: i) an explicit description of quantum mechanical
histories defined on the classical phase space, and ii) the
translation of the usual methods of perturbation theory in the
history context. In both issues, our emphasis lies on the
construction of the decoherence functional, from which all physical
predictions of the theory (probabilities) are derived. These results
then allow the translation of common and useful techniques of
standard quantum mechanics in the histories framework and provide
therefore a tool for addressing problems of a technically more
complex nature.

A history is represented by  a time-ordered string of projection
operators representing propositions about the system. We denote
these $\a:=(\a_{t_1},\a_{t_2},\ld,\a_{t_n})$ with $t_1<t_2<\ld<t_n$.
Given a Hamiltonian $H$, and an initial state $\r_0$, the
decoherence functional is defined on pairs of histories as: \be
\label{eq:df} d(\a,\b)=Tr[C^{\dag}_\a\r_0C_\b]\ee in which the class
operator is given in terms of Heisenberg picture projection
operators as $C_\a=\a_{t_1}(t_1)\a_{t_2}(t_2)\ld\a_{t_n}(t_n)$.

Over the last twenty years, the histories formalism has undergone
significant developments. For example, the `history projection
operator' (HPO) approach of Isham, Linden and Savvidou
\cite{HPOa},\cite{Sav98},\cite{Sav99a},\cite{Sav01a} (which
heavily influences the current work) has focussed on the
temporo-logical structure of histories, whereas Hartle's
`generalized quantum mechanics' \cite{Har93} and Savvidou's
spacetime description of HPO histories
\cite{Sav01a},\cite{SavGR03} (see also \cite{Bur03}) cast quantum
theory into a more manifestly covariant form.

In the present work we are interested in the further development
of the phase space description of quantum mechanical histories
initiated by Anastopoulos \cite{Ana00a},\cite{Ana02} and the
development of a generic method for the implementation of
perturbative techniques in the histories formalism. Central to
this work is the identification of the decoherence functional with
the closed-time-path (CTP) generating functional (first introduced
by Schwinger \cite{Sch61}). From this we gain both a definition of
the decoherence functional via a quasi-distribution on the
classical histories phase space $\Pi$ (this is the space of all
continuous paths on the standard phase space---see \cite{Sav99a},
Ch. 5), and, as the latter has a well-defined perturbative
expansion, a starting point for the description of interacting
theories in the histories formalism.

The paper is structured as follows. We briefly present the
necessary background material in Section \ref{sec:background2},
largely to fix notation. In Section \ref{sec:psrqmh}, we perform
an explicit analysis of different phase space representations of
the decoherence functional for the case of the simple harmonic
oscillator (SHO). We construct a parameterised expression for the
phase space distribution, given a generic initial state, at the
discrete-time level which allows us to compare the most commonly
encountered representations, ie. Q, P and Wigner. Concluding that
the P representation is ill-defined, we then take the
continuous-time limit of the coherent state (Q) and Wigner
representations. Contrary to the single-time case, we find
significant differences in the structure of these representations
in the context of histories. In Section \ref{sec:dfaho}, we make
use of the Fourier transform relationship between the CTP
generating functional and the (quasi) distribution to develop, at
the continuous-time level, a perturbative expression for the
decoherence functional of the quartic anharmonic oscillator (AHO).
We construct this expression to second order in the coupling
constant in both the configuration space and phase space contexts.
In Section \ref{sec:conclusion} we conclude.

\section{Background}\label{sec:background2}
\subsection{Quantum Mechanical Phase Space
Representations}\label{sec:stpsr}

Following \cite{AW70a}, we associate a c-number function on the
classical phase space $\Ga$, with an operator
$G(\hat{a}^{\dag},\hat{a})$ that is a function of the non-commuting
operators $\hat{a}^{\dag}$ and $\hat{a}$, on the Hilbert space $\H$,
according to: \bea \label{eq:c-q rep}
F^{\Om}_G(\a,\a^*)\!\!\!\!\!&=&\!\!\!\!\!Tr[G(\hat{a}^{\dag},\hat{a})
\De^{\tilde{\Om}}(\a,\a^*)]\\
\!\!\!\!\!&=&\!\!\!\!\!\int\!\!\! d^2\!z\!~\exp\!\{-z
\a\!^*\!+\!z\!^*\a\}\tilde{\Om}(z,z\!^*\!)Tr[G(\hat{a}^{\dag},
\hat{a})\exp\{\hat{a}^{\dag}z-\! \hat{a}z\!^*\},]\eea where
$\hat{a}^{\dag}$ and $\hat{a}$ are the boson creation and
annihilation operators and where $d^2z$ is the standard Lebesque
measure normalised by $2\pi$. $\Om(z,z^*)$ represents the linear
mapping $\hat{G}=\Om\{F\}$ and
$\tilde{\Om}(z,z^*)=[\Om(-z,-z^*)]^{-1}$. $\Om(z,z^*)$ is intimately
related to the ordering of non-commuting operators.

The most commonly encountered phase space representations are
members of the sub-class of mappings given simply by
\be\label{eq:spsm}
\Om\{z,z^*\}=\exp\big{\{}\frac{s}{2}|z|^2\big{\}}.\ee For $s=1,0,-1$
we have, respectively, the Q, Wigner and P representations. The
phase space representation of the density operator is often referred
to as a quasi-probability distribution as it plays an analogous role
to a classical probability distribution, ie:\be\label{eq:TrrhoB}
Tr[\r G(\hat{a}\!^{\dag}\!,\!\hat{a})]\!\!\equiv\!\!\langle
G(\hat{a}\!^{\dag}\!,\!\hat{a})\rangle_\r\!\!=\!\!\!\int\!\!
d^2\!zF_\r^{\Om}(z,\!z\!^*\!)F_G^{\tilde{\Om}}(z,z\!^*\!)=\!\!\!\int\!\!
d^2\!zF_\r^{\tilde{\Om}}(z,\!z\!^*\!)F_G^{\Om}(z,\!z\!^*\!).\ee

In \cite{Sri77}, Srinivas describes the generalisation of these
results so as to write multi-time quantum correlation functions in
terms of a multi-time quasi-probability distribution on phase space.
The latter is given by:\be\label{eq:mtqpd}
F^{\Om}_\r(\a_1,\a_1^*,t_1;\ld ;\a_n,\a_n^*,t_n)\equiv Tr[\r_0
\De^{\tilde{\Om}}(\a_1,\a_1^*,t_1)\ld
\De^{\tilde{\Om}}(\a_n,\a_n^*,t_n)],\ee in which the time-evolved
representation operator is given by
$\De^{\Om}(\a,\a^*,t)=e^{itH}\De^{\Om}(\a,\a^*)e^{-itH}$.

\subsection{The Relationship Between the CTP Generating Functional
and the Decoherence Functional}\label{sec:ctpdf}

The reader is referred to \cite{Ana00a} for a detailed account of
what follows.

The decoherence functional, Eq. (\ref{eq:df}), defined on general
operators, can be understood as the expectation value of \emph{two}
strings of operators---one time-ordered and one anti-time-ordered.
These correlation functions are generated by the CTP generating
functional. Thus, if the history Hilbert space $\V$ (defined as a
tensor product of copies of the standard Hilbert space), carries a
representation of the history Weyl group
$U(\x(\cd),\c(\cd))=\exp\{iq_\x+ip_\c\}$ (in which $q_\x$ and $p_\c$
are the time-averaged position and momentum operators) then we can
define the configuration space CTP generating functional as
$Z[\x,\x']=d(e^{iq_\x},e^{iq_\x'})$. Furthermore, we can construct
the phase space CTP generating functional which will contain all the
physical information about the system. This will be given by
\be\label{eq:phaseCTP} Z[\x,\c;\x',\c']=d(U(\x,\c),U(\x',\c')).\ee
Note that the CTP generating functional thus inherits the
normalisation condition $Z[0,0]=1$.

Denoting for simplicity a phase space path $\g\equiv[q(\cd),
p(\cd)]$, we can formally associate to the decoherence functional a
quasi distribution $W^{\Om}[\g|\g']$, on the histories phase space
(strictly speaking on $\Pi\times\Pi$) according to \be
\label{eq:ddd}
d(A,B)=\int\D\g\D\g'\,W^{\Om}[\g|\g']F^{\tilde{\Om}}_A[\g]F^{
\tilde{\Om}}_B[\g'],\ee in which $A$ and $B$ are operators on $\V$,
and $F^{\tilde{\Om}}_A[\g]$ is a phase space representation of $A$
defined in analogy to Eq. (\ref{eq:c-q rep}).

The phase space CTP generating functional is related to the
continuous-time histories phase space quasi-distribution by a
(functional) Fourier transform
$W^{\Om}[q(\cd),p(\cd)|q'(\cd),p'(\cd)]=\int\D\x\D\c\D\x'\D\c'~e^{iq\cd\x+ip\cd\c
-iq'\cd\x'-ip'\cd\c'}Z^{\Om}[\x(\cd),\c(\cd)|\x'(\cd),\c'(\cd)]$, in
which we have used the shorthand $q\cd\x\equiv\int dt\,q(t)\x(t)$.

The integration over paths in the above expressions is formal. It
is properly defined by a consideration of discrete-time histories,
the definition of suitable cylinder sets in the space of
continuous time histories and extension by continuity to a larger
class of phase space paths (see \cite{Ana00a} for proof).

\section{Phase Space Representations for the SHO}\label{sec:psrqmh}

In this section, we explicitly construct and analyse the different
phase space representations of the decoherence functional for the
case of a single harmonic oscillator, described by the Hamiltonian
\be \hat{H}~=~:\fr
\hat{p}^2+\fr\o^2\hat{q}^2:~=~\o\hat{a}^{\dag}\hat{a}.\ee The key
new results of this section are (i) the computation of the
discrete-time expression for the distribution, parameterised
according to Eq. (\ref{eq:spsm}), given a generic initial state,
(ii) explicit expressions for the Wigner and Q representations at
both the discrete-time and continuous-time level, and the infinite
time ($t\in\mathbb{R}$) limit of the latter, and (iii) the form of
the CTP generating functional that arises in this limit. (This last
expression will form the basis of the development of the
perturbation theory of the AHO in the next section.)

We employ the following relations to interchange between complex and
real coordinates on $\Ga$ and its dual \bea \a &=& \sqrt{\o/2}q + i
1/ \sqrt{2 \o} p \\ z &=& - \sqrt{\o/2}\c + i 1/ \sqrt{2 \o} \x.\eea

The continuous-time phase space distribution $W^{\Om}[\g|\g']$ is
defined as the limit of the following discrete-time expression
\cite{Ana00a} \be\label{eq:mtqpd0}
W^{\Om}\!_{n,m}(\a\!_1,\a\!^*\!_1\!,\!t_1\!;\!\ld\!;\!\a\!_n,\!\a\!^*\!_n,\!t\!_n|\a'_1,\a^{\prime*}_1,t'_1;
\ld;\a'_m,a^{\prime*}_m,t'_m)=Tr(\hat{C}^{\dag}_n\r_0\hat{C}_m),\ee
in which $\hat{C}_n=\De^{\tilde{\Om}}(\a_1,\a^*_1,t_1)\ld
\De^{\tilde{\Om}}(\a_n,\a^*_n,t_n)$. The `branches' are time-ordered
so that $t_1<t_2<\ld<t_n$ and $t'_1<t'_2<\ld<t'_m$. This is the
histories theory generalisation of Eq. (\ref{eq:mtqpd}).

As the expressions are somewhat unwieldy in full, we only
demonstrate the calculation on one `branch', ie:
\bea\label{eq:mtqpd2}
\lefteqn{W^{\Om}_n(\a_1,\a^*_1,t_1;\ld;\a_n,\a^*_n,t_n)\equiv
Tr(\r_0\hat{C}_n)=}\nn\\&& \int d^2z_1\ld d^2z_n
\,e^{-\sum_{i=1}^n(\a^*_iz_i-\a_iz^*_i)}e^{-\frac{s}{2}\sum_{i=1}^n|z_i|^2}
\times\nn\\&&Tr\left[\r_0 e^{it_1H}U(z_1,z^*_1)e^{-it_1H}\ld
e^{it_nH}U(z_n,z^*_n)e^{-it_nH}\right],~~~~~\eea in which the Weyl
operators are given by
$U(z,z^*)=\exp\{\hat{a}^{\dag}z-\hat{a}z^*\}$.

To maintain generality as to the initial state, it suffices to
choose a coherent state as all density matrices can be written as a
weighted, diagonal sum of such states. Thus we take
$\r_0=\ket{\b}\bra{\b}$, in which
$\ket{\b}:=e^{\hat{a}^{\dag}\b-\hat{a}\b^*}\ket{0}$ for
$\b\in\mathbb{C}$.

Using the composition law \be\label{eq:Weylcomp}
U(z,z^*)U(z',z^{\prime
*})=e^{\fr(z^{\prime*}z-z^*z')}U(z+z',z^*+z^{\prime*})\ee and the
time evolution $e^{iHt}U(z,z^*)e^{-iHt}=U(e^{i\o t}z,e^{-i\o
t}z^*)$, the result is
\bea\label{eq:mtqpd3}\lefteqn{W^{\Om}_n(\a_1,\a^*_1,t_1;\ld;\a_n,\a^*_n,t_n)
=\exp\Big{\{}\frac{4}{1+s}\sum_{i,j=1}^n
\th(t_j-t_i)A_{j-i}(s)e^{-i\o(t_i-t_j)}\a^*_i\a_j\Big{\}}\times}\nn\\&&
\exp\Big{\{}-2\b^*\b\sum_{i=1}^nA_i(s)+2\b\sum_{i=1}^n\a^*_iA_{n+1-i}(s)e^{-i\o
t_i}+ 2\b^*\sum_{i=1}^n\a_iA_i(s)e^{i\o t_i}\Big{\}}~~~~~~~~\eea
in which the coefficients $A_k(s)$ are given by:\bea
A_0(s)&=&-1\\A_1(s)&=&\frac{1}{1+s}\\A_{k+1}(s)&=&\left[1-2\left(\frac{1}{1+s}
\right)\right]A_k(s)~~~(k\geq 1)\eea  and in which the step
function is given by:\be\th(t_i-t_j)
=\left\{\begin{array}{ll} 1& t_i>t_j\\
\fr & t_i=t_j\\ 0 & t_i<t_j\end{array}\right.\ee The second
exponential on the right hand side of Eq. (\ref{eq:mtqpd3}) contains
the boundary terms that arise as the result of our choice of initial
state.

The result of the full calculation of  Eq. (\ref{eq:mtqpd0})
contains a similar expression for the other `branch' (with opposite
time-ordering) and an expression involving cross-terms between the
primed and unprimed quantities. It is given by:

\bea\label{eq:mtqpd4}\lefteqn{W^{\Om}_{n,m}(\a_1,\a^*_1,t_1;\ld;\a_n,
\a^*_n,t_n|\a'_1,\a^{\prime*}_1,t'_1;\ld;\a'_m,\a^{\prime*}_m,t'_m)=}
\nn\\&&
\exp\Big{\{}\frac{4}{1+s}\sum_{i,j=1}^n\th(t_i-t_j)A_{i-j}(s)e^{-i
\o(t_i-t_j)}\a_i\a^*_j\Big{\}}\times\nn\\&&
\exp\Big{\{}\frac{4}{1+s}\sum_{i,j=1}^m\th(t'_j-t'_i)A_{j-i}(s)
e^{-i\o(t'_i-t'_j)}\a^{\prime*}_i\a'_j\Big{\}}\times\nn\\&&
\exp\Big{\{}4\sum_{i=1}^n\sum_{j=1}^mA_{n+1-i}(s)A_{m+1-j}(s)
e^{-i\o(t'_j-t_i)}\a_i\a^{\prime*}_j\Big{\}}\times\nn\\&&
\exp\Big{\{}-2\b^*\b\sum_{i=1}^nA_i(s)+2\b\sum_{i=1}^n\a^*_iA_i(s)e^{-i\o
t_i}+2\b^*\sum_{i=1}^n\a_iA_{n+1-i}(s)e^{i\o
t_i}\Big{\}}\times\nn\\&&
\exp\Big{\{}-2\b^*\b\sum_{i=1}^mA_i(s)+2\b\sum_{i=1}^m\a^{\prime*}_iA_{m+1-i}
(s)e^{-i\o t'_i}+2\b^*\sum_{i=1}^m\a'_iA_i(s)e^{i\o
t'_i}\Big{\}}\times\nn\\&&
\exp\Big{\{}4\b^*\b\sum_{i=1}^n\sum_{j=1}^mA_i(s)A_j(s)-4\b\sum_{i=1}^n
\sum_{j=1}^m\a^{\prime*}_jA_{m+1-j}(s)A_i(s)e^{-i\o t'_j}-\nn\\&&
4\b^*\sum_{i=1}^n\sum_{j=1}^m\a_iA_j(s)A_{n+1-i}(s)e^{i\o
t_i}\Big{\}}.~~~~~\eea This is the full, discrete-time expression
for the phase space distribution associated to the decoherence
functional of the SHO with a generic initial state by a general
rule of association given by
$\Om(z,z^*)=\exp\big{\{}\frac{s}{2}|z|^2\big{\}}$. This includes
the Q, P and Wigner representations for $s=1,-1,0$ respectively.

It is clear that the expression Eq. (\ref{eq:mtqpd4}), is not
well-defined for $s=-1$, as the first two exponents have $1+s$ in
the denominator, which introduces an infinity into the expression
for the distribution. Thus we conclude that the P representation is
not a good choice for representing quantum mechanical histories on
the classical histories phase space. We shall now examine, in turn,
the Q representation ($s=1$) and the Wigner representation ($s=0$),
and their respective continuum limits.

\subsection{The Q representation}\label{sec:cthpsrQ}

The Q representation is given by $s=1$, and thus $A_k(s)=0$ for
$k\geq 2$. The resulting expression is thus local in time. This is
an important property, as it transpires that the phase space
decoherence functional satisfies a histories version of the Markov
property \cite{Ana02}.

Taking $s=1$, Eq. (\ref{eq:mtqpd3}) becomes
\bea\label{eq:mtqpdq}\lefteqn{W^Q_n(\a_1,\a^*_1,t_1;\ld
;\a_n,\a^*_n,t_n)=\exp\Big{\{}-\b^*\b+\b\a_n^*e^{-i\o
t_n}+\b^*\a_1e^{i\o
t_1}\Big{\}}}\nn\\&&\times\exp\Big{\{}-|\a_n|^2+\sum_{i=1}^{n-1}
\left(-|\a_i|^2+\a^*_i\a_{i+1}e^{-i\o(t_i-t_{i+1})}\right)\Big{\}}
.\eea To get the continuous-time limit, we take $n$ large (so
$t_{i+1}-t_i\equiv\d t<<1$), define the discrete derivative
$\dot{\a}_i=\frac{\a_{i+1}-\a_i}{\d t}$, neglect terms of $O(\d
t^2)$, and then we take $\d t\rightarrow 0$. The result is \bea
\lefteqn{W^Q[\a(\cd),\a^*(\cd)]=}\nn\\&&\exp\Big{\{}-|\b|^2+
\b\a_n^*(t_n)+\b^*\a_1-|\a_n|^2+\int_{t_1}^{t_n}dt~\Big{(}\a^*(t)
\dot{\a}(t)+i\o\a^*(t)\a(t)\Big{)}\Big{\}}.~~~~~~~\eea The result
for the full expression, i.e. the continuum limit of Eq.
(\ref{eq:mtqpd0}) in the Q representation is \bea\label{eq:ctpsdq}
\lefteqn{W^Q[\a(\cd),\a^*(\cd)|\a'(\cd),\a^{\prime*}(\cd)]=
\exp\Big{\{}-|\b|^2+\b\a_1+\b^*\a'_1+\a_m^{\prime*}\a_n+}\nn\\&&
\int_{t_1}^{t_n}dt~\Big{(}\dot{\a}^*(t)\a(t)-i\o\a^*(t)\a(t)\Big{)}+
\int_{t'_1}^{t'_m}dt~\Big{(}\a^{\prime*}(t)\dot{\a}'(t)+i\o\a^{\prime*}
(t)\a'(t)\Big{)}\Big{\}}~~~~~~~~\eea in which we have also taken
$t_n=t'_m$ as is usual in histories.

In the infinite-time limit, ie.
$[t_1,t_n],[t'_1,t'_m]\rightarrow[-\infty,\infty]$, the
requirement of square-integrability on the space of paths forces
the single-time Hilbert spaces at $t=\pm\infty$ to be
one-dimensional, consisting only of the vector $\ket{0}$. Thus the
initial state is the vacuum, ie. $\r_0=\ket{0}\bra{0}$. If we
revert to the $[q(\cd),p(\cd)]$ coordinates on $\Pi$, the above
expression, in this limit, becomes simply \be\label{eq:ctpsdq1}
W^Q[q(\cd),p(\cd)|q'(\cd),p'(\cd)]=\exp\Big{\{}iS[q(t),p(t)]-iS[q'(t),
p'(t)]\Big{\}},\ee where $S[q(t),p(t)]=\int
dt~\left[p(t)\dot{q}(t)-\fr\big{(}p(t)^2+\o^2q(t)^2\big{)}\right]$
is the phase space action functional.

\subsection{The Wigner representation}\label{sec:cthpsrW}

In the case of the Wigner representation, $s=0$, we get a
non-local expression with $A_{k+1}(s)=-A_k(s)$. For one `branch'
we have
\bea\label{eq:mtqpdw}\lefteqn{W^{Wigner}_n(\a_1,\a^*_1,t_1;\ld
;\a_n,\a^*_n,t_n)=}\nn\\&&\exp\Big{\{}4\sum_{i,j=1}^n
\th(t_j-t_i)(-1)^{i+j+1}e^{-i\o(t_i-t_j)}\a^*_i\a_j\Big{\}}\times\nn\\&&
\exp\Big{\{}-2\b^*\b\sum_{i=1}^n(-1)^{i+1}+2\b\sum_{i=1}^n(-1)^{n-i}
\a^*_ie^{-i\o t_i}+2\b^*\sum_{i=1}^n(-1)^{i+1}\a_ie^{i\o
t_i}\Big{\}}.~~~~~~~~\eea The alternating sign makes the
calculation of the continuum limit of this expression a little
more tricky. However, a similar situation was encountered in
\cite{SA06}, and a solution detailed in Section III.C and Appendix
B of that reference. The only difference is that the current
expressions contain an explicit time dependence, however this does
not complicate the derivation in any significant way. We outline
the calculation of the continuum limit of the above expression in
the Appendix. The final result is \bea\label{eq:ctqpdw}
\lefteqn{W^{Wigner}[\a(\cd),\a^*(\cd)|\a'(\cd),\a^{\prime*}(\cd)]=}
\nn\\&&\exp\Big{\{}\fr|\a_1|^2-\fr|\a_n|^2-\a_1(t_1)\a^*_n(t_n)-
\int_{t_1}^{t_n}dt~\big{(}\a^*(t)\dot{\a}(t)+i\o\a^*(t)\a(t)\big{)}\Big{\}}\times\nn\\&&
\exp\Big{\{}\fr|\a'_1|^2-\fr|\a'_m|^2-\a^{\prime*}_1(t'_1)\a'_m(t'_m)-
\int_{t'_1}^{t'_m}dt~\big{(}\a'(t)\dot{\a}^{\prime*}(t)-i\o\a^{\prime*}(t)\a'(t)\big{)}\Big{\}}\times\nn\\&&
\exp\Big{\{}\a_n\a^{\prime*}_m-\a_n(t_n)\a^{\prime*}_1(t'_1)-
\a_1(t_1)\a^{\prime*}_m(t'_m)-\a_1(t_1)\a^{\prime*}_1(t'_1)\Big{\}}\times\nn\\&&
\exp\Big{\{}\b\big{(}\a^{\prime*}_m(t'_m)-\a^{\prime*}_1(t'_1)-
\a^*_n(t_n)+\a^*_1(t_1)\big{)}+\nn\\&&\b^*\big{(}\a_n(t_n)-\a_1(t_1)-
\a'_m(t'_m)+\a'_1(t'_1)\big{)}\Big{\}}.~~~~~~~~\eea

Thus, in the continuum limit, the classical action reappears along
with a collection of boundary terms. These latter are
significantly different from those that appear in the Q
representation and reflect the symmetric nature of the Wigner
representation.

Taking the infinite-time limit again, we find that this result
reduces to the same expression as we obtained for the Q
representation, namely \be
W^{Wigner}[q(\cd),p(\cd)|q'(\cd),p'(\cd)]=\exp\Big{\{}iS[q(t),p(t)]-iS[q'(t)
,p'(t)]\Big{\}}.\ee This is a satisfying result as the classical
limits of these expressions should be the same. However, we have
compared the Q and Wigner representations in the far more general
context of a generic initial state, at both the
discrete-time---Eqs. (\ref{eq:mtqpdq}) and (\ref{eq:mtqpdw})
respectively---and continuous-time---Eqs. (\ref{eq:ctpsdq}) and
(\ref{eq:ctqpdw})---levels.

The discrete-time expressions are the more fundamental as it is at
this level that one may define the generic coarse-graining
operations that are central to the histories formalism. We have seen
the difference in the boundary terms that arise in each
representation. But, most importantly, we get a local expression for
the Q representation. As we mentioned earlier, this allows us the
definition of a Markov property for the distribution and thus the
decoherence functional. This is in line with the properties of the
wave function (or density matrix) in standard QM, and for this
reason we conclude that the Q representation is the most suitable
for discussing the phase space structure of the histories formalism.

This analysis complements the discussion in \cite{Ana02} in which it
is suggested that a representation based on coherent states is more
suitable, and where it was shown that a Markov property for the
phase space decoherence functional is a necessary but not sufficient
condition for a reconstruction theorem, which regains the standard
Hilbert space from the phase space picture.

\subsection{The phase space CTP generating functional}\label{sec:cfs}

In this section we compute the phase space CTP generating functional
in the infinite time limit. This is given by the Fourier transform
of the distribution. It will also form the basis of the perturbative
expressions calculated in the next section. Finally, we calculate
and briefly discuss the correlation functions that arise.

Since the Q and Wigner representations coincide in this limit,
namely \be
W^Q[q(\cd),p(\cd)|q'(\cd),p'(\cd)]=W^{Wigner}[q(\cd),p(\cd)|q'(\cd),p'(\cd)]=\exp\Big{\{}iS[q(t),p(t)]-iS[q'(t)
,p'(t)]\Big{\}}\ee we drop the superscript in what follows, however,
it should be remembered that, in more general situations, the Q
representation is the more appropriate. The CTP generating
functional will be given by \bea\label{eq:ctpFT}
\lefteqn{Z[\x(\cd),\c(\cd)|\x'(\cd),\c'(\cd)]=}\nn\\&&\int\D q\D p\D
q'\D
p'~e^{-iq\cd\x-ip\cd\c+iq'\cd\x'+ip'\cd\c'}W[q(\cd),p(\cd)|q'(\cd),p'
(\cd)]\\&&=\exp\Big{\{}-\fri\int
dtdt'\,\Big{[}\big{(}\x(t)-\dot{\c}(t)\big{)}\De_F(t-t')\big{(}\x(t')
-\dot{\c}(t')\big{)}\nn\\&&+2\big{(}\x(t)-\dot{\c}(t)\big{)}\De^+(t-t')
\big{(}\x'(t')-\dot{\c}'(t')\big{)}\nn\\&&
-\big{(}\x'(t)-\dot{\c}'(t)\big{)}\De_D(t-t')\big{(}\x'(t')-\dot{\c}'
(t')\big{)}\Big{]}\nn\\&&+\fri\int
dt\,\left(\c(t)^2-\c'(t)^2\right)\Big{\}}.\eea The Green's functions
are \bea\label{eq:Feynmanq}
\De_F(t-t')&=&-i\bra{0}T[q(t)q(t')]\ket{0}=\int\frac{dk}{2\pi}\frac{1}
{k^2-\o^2+i\ep}\,e^{-ik(t-t')}\\
\De^+(t-t')&=&i\bra{0}q(t)q(t')\ket{0}=\frac{1}{2\o}e^{i\o(t-t')}\\
\De_D(t-t')&=&i\bra{0}\tilde{T}[q(t)q(t')]\ket{0}=\int\frac{dk}{2\pi}
\frac{1}{k^2-\o^2-i\ep}\,e^{-ik(t-t')},\label{eq:Dysonq}\eea where
$\tilde{T}$ indicates \emph{anti}-time ordering. These are,
respectively, the Feynman, Wightman and Dyson Green's functions.

By writing $X^a(a=1,2)\equiv(q,p)$ and $J^a(a=1,2)\equiv(\x,\c)$, we
have the general expression for a mixed $(n,m)$ correlation function
\bea \lefteqn{\G^{n,m}(a_1,t_1;\ld;a_n,t_n|b_1,t'_1;\ld;b_m,t'_m)=}
\nn\\&&(-i)^ni^m\frac{\d^{n+m}}{\d J^{a_1}(t_1)\ld \d J^{a_n}(t_n)\d
J^{'b_1}(t'_1)\ld \d
J^{'b_m}(t'_m)}Z[\x,\c,|\x',\c']\Big{|}_{\x=\c=\x'=\c'=0}.\eea

The two-point functions are readily computed from this expression.
Generally, they are as would be expected from standard QM \emph{with
the exception of the following} \bea
\bra{0}T[p(t)p(t')]\ket{0}&=&i\pa_t\pa_{t'}\De_F(t-t')-i\d(t-t')
\label{eq:Tpp}\\
\bra{0}\tilde{T}[p(t)p(t')]\ket{0}&=&-i\pa_t\pa_{t'}\De_D(t-t')+i\d(t-t')
\label{eq:T'pp} \eea (we have abbreviated $\pa/\pa t\equiv \pa_t$).
Though we will not discuss these in detail here, these reflect the
fact that, in histories theories, we do not have $p(t)=\dot{q}(t)$
in general. Most relevant to the discussion here is the analysis in
\cite{Ana02} in which the difference between velocity and momentum
is determined by a `random external force'---which arises in the
quantum analogue of a stochastic differential equation.

In histories---as opposed to single-time---quantum theory, we can
define differentiation with respect to time independently of the
dynamical evolution. This leads to the definition of a velocity
operator that is independent of, and does not generally commute with
the momentum operator \cite{Sav98}. In \cite{AS06}, probabilities
for measurements that are extended in time are considered; class
operators are constructed that are significantly different for
momentum and velocity measurements\footnote{although they coincide
for large coarse-graining} and the scheme suggests that it may be
possible to experimentally distinguish between the two.

\section{The Decoherence Functional for the AHO}\label{sec:dfaho}

In this section we develop a perturbative method to extend the
formalism of continuous-time histories to include interacting
theories. We exemplify the construction in the case of the
anharmonic oscillator with a quartic self-interaction,
corresponding to a Hamiltonian \be\label{AHOhamiltonian}
H(p,q)=\fr p^2+\fr\o^2q^2+\frac{\l}{4!}q^4.\ee The key to this
development is the relationship between the decoherence functional
and the CTP generating functional described in Section
\ref{sec:ctpdf}. Though we have stressed that the discrete-time
expressions for the decoherence functional are to be considered
the more fundamental, the role played by the CTP generating
functional allows us to import the standard mathematical
techniques of perturbation theory, for which the underlying
discrete-time expressions, and the path integral expressions that
arise in the continuous-time limit are well-established. Thus we
can, at this stage, work confidently with the functional
expressions. Of course to be fully rigorous, one would want to
start from the discrete-time expressions - say a discrete-time
version of Eq. \ref{eq:intcsCTP} below - but we feel that the
results presented here are sufficiently rigorous for the purpose
of this paper.

We work both in configuration space and phase space, and the main
results of this section are the calculation of a perturbative
expression for the distribution up to second order in the coupling
constant in each case. This method can readily be extended to
encompass more physically realistic situations as the functional
techniques that we have adopted are well-defined for initial
states other than the vacuum. We also briefly discuss the
relationship between the decoherence functional and the CTP
effective action.

From here on, a subscript `$0$' will indicate a quantity referring
to the free theory, ie. for $\l=0$.

\subsection{The perturbative expansion on configuration
space}\label{sec:pecs}

The construction of the distribution and CTP generating functional
for the SHO on configuration space is straightforward as we face no
complications with non-commuting operators. The latter is simply
derived from consideration of the two-point functions and is given
by (see, eg. \cite{CH86}) \bea\label{eq:freecsCTP}
Z_0[J,J']&=&\exp\Big{\{}-\fri\int dtdt'
\Big{(}J(t)\De_F(t-t')J(t')+\nn\\&&
2J(t)\De^+(t-t')J'(t')-J'(t)\De_D(t-t')J'(t')\Big{)}\Big{\}}.~~~~~
\eea The Green's functions were given in Eqs.
(\ref{eq:Feynmanq})-(\ref{eq:Dysonq}), and we note that the Wightman
Green's function is a solution to the SHO equation of motion.

The configuration space distribution will be given by the Fourier
transform of this expression, i.e., \be d_0[q(\cd),q'(\cd)]=\int\D
J\D J'~e^{iq\cd J-iq'\cd J'}Z_0[J,J']\ee and this is readily
calculated to be \be\label{eq:freecsDF}
d_0[q(\cd),q'(\cd)]=e^{iS[q]-iS^*[q']}\ee where $S[q]$ is the
classical action calculated on the path $q(t)$ \be S[q]=\fr\int
dt~(\dot{q}(t)^2-\o^2q(t)^2)=\fr\int
dtdt'~q(t)\De_F(t-t')^{-1}q(t').\ee The complex conjugation reflects
the anti-time ordering on the $q'$ path. This result is exactly as
we would expect. It is the decoherence functional defined on a pair
of fine-grained configuration space paths (see, eg. \cite{Har93}
where the same result is derived by considering class operators made
up of strings of Heisenberg picture projection operators onto
regions of configuration space).

Given a family of commuting, self-adjoint operators $\hat{A}^i$,
subsets of $\Pi_{CS}$---the space of configuration space
histories---will correspond to histories of the observables
$\hat{A}^i$. If we consider two such subsets $C$ and $D$, we can
formally write the configuration space decoherence functional
\cite{Ana02}
 \be\label{eq:csDF} d(C,D)=\int \D q\D
q'~d[q(\cd),q'(\cd)]\c_C[q(\cd)]\c_D[q'(\cd)]\ee where $\c_C$ is the
characteristic function associated with the subset
$C\subset\Pi_{CS}$.

In the case of the AHO with a quartic self-interaction, the CTP
generating functional will be given by \be\label{eq:intcsCTP}
Z[J,J']=\N \exp\left\{-\frac{i\l}{4!}\int dt\left(\frac{\d^4}{\d
J(t)^4}-\frac{\d^4}{\d J'(t)^4}\right)\right\}Z_0[J,J'].\ee The
normalisation condition described in Section \ref{sec:ctpdf} above,
i.e., $Z[0,0]=1$, is equivalent to the cancelling of vacuum diagrams
from this series.

The configuration space distribution for the interacting theory is
given by the Fourier transform of $Z[J,J']$, ie:
\be\label{eq:csFT} d[q(\cd),q'(\cd)]=\int\D J\D J'~e^{iq\cd
J-iq'\cd J'}Z[J,J'].\ee The result is most clearly expressed in an
exponential form, ie.
$d[q(\cd),q'(\cd)]=e^{i\tilde{S}[q(\cd),q'(\cd)]}$. (This is
analogous to the situation in standard field theory when one works
with the generating functional for the connected Green's
functions, $W[J]=-i\ln Z[J]$.) After some lengthy, but relatively
straightforward calculation, we arrive at the following
perturbative expression for $\tilde{S}[q(\cd),q'(\cd)]$:\bea
\lefteqn{i\tilde{S}[q(\cd),q'(\cd)]=i\tilde{S}_0[q(\cd),q'(\cd)]-\frac{i\l}{4!}\int
dt\,\Big{(}q(t)^4-q'(t)^4\Big{)}}\nn\\&&
+\fr\left(-\frac{i\l}{4!}\right)^2\int dtdt'\,\Big{(}
192iq(t)\De^+(t-t')^3q'(t')+144q(t)^2\De^+(t-t')^2q'(t')^2\nn\\&&-32iq(t)
^3\De^+(t-t')q'(t')^3\Big{)}+0(\l^3),\eea in which
$\tilde{S}_0[q(\cd),q'(\cd)]=S[q]-S^*[q']$. This is the main
result of this part of the current work, and in the next section
we shall see that the result for the phase space distribution is
essentially the same. These expressions can then be used in Eq.
(\ref{eq:csDF}) (for configuration space) and Eq. (\ref{eq:ddd})
(for phase space), along with suitable coarse-grainings, in order
to determine a perturbative expression for the decoherence
functional of coarse-grained histories of the AHO.

Writing the decoherence functional in this manner raises the
interesting question of how $\tilde{S}[q,q']$ is related to the
CTP effective action. The latter is defined by a double Legendre
transform of the generating functional of connected diagrams
$W[J,J']=-i\ln Z[J,J']$  \be
\Ga_{CTP}[\bar{q},\bar{q}']=W[J,J']-J\cd \bar{q}+J'\cd
\bar{q}',\ee where the sources $(J,J')$ are considered as
functionals of the background fields $(\bar{q},\bar{q}')$, which
are, in turn, defined as $\bar{q}=\frac{\pa W[J,J']}{\pa J}$,
$\bar{q}'=-\frac{\pa W[J,J']}{\pa J'}$ \cite{CH86}.

In \cite{CH94} it is conjectured that the decoherence functional
is defined by the tree-level CTP effective action
$d[q,q']=e^{i\Ga_{CTP}[q,q']}$. We can now show that, to $O(\l)$
at least, this is indeed the case, as the CTP effective action for
the SHO is given by \bea
\lefteqn{\Ga_{CTP}[q,q']=\tilde{S}_0(q,q')-}\nn\\&&
\frac{\l}{4!}\int
dt~[q(t)^4-q'(t)^4+\De_F(0)q(t)^2+\De_D(0)q'(t)^2]+O(\l^2)\eea and
we ignore the last two, `one-loop' terms.

\subsection{The perturbative expansion on phase
space}\label{sec:peps}

In Section \ref{sec:psrqmh} we discussed in some detail the
construction of the phase space distribution for continuous-time
histories. In the infinite time limit, the Q and Wigner
representations coincided, given by \be
W_0[q(\cd),p(\cd)|q'(\cd),p'(\cd)]=\exp\Big{\{}iS[q(t),p(t)]-iS[q'(t)
,p'(t)]\Big{\}}.\ee The corresponding CTP generating functional was
given in Section \ref{sec:cfs}. The phase space CTP generating
functional for the interacting theory will be given by \bea
\lefteqn{Z[\x(\cd),\c(\cd)|\x'(\cd),\c'(\cd)]=}\nn\\&&\N
\exp\left\{-\frac{i\l}{4!}\int dt\left(\frac{\d^4}{\d
\x(t)^4}-\frac{\d^4}{\d
\x'(t)^4}\right)\right\}Z_0[\x(\cd),\c(\cd)|\x'(\cd),\c'(\cd)]~~~~~
\eea and, once more, the distribution will be given by the Fourier
transform of the resulting expression, \bea\label{eq:ctpFT1}
\lefteqn{W[q(\cd),p(\cd)|q'(\cd),p'(\cd)]=}\nn\\&&
\int\D\x\D\c\D\x'\D\c'~e^{iq\cd\x+ip\cd\c-iq'\cd\x'-ip'\cd\c'}Z[\x(\cd)
,\c(\cd)|\x'(\cd),\c'(\cd)].\eea

These calculations are greatly simplified if we transform to new
coordinates $u(t)=\x(t)-\dot{\c}(t)$ and $v(t)=\c(t)$, and
likewise for the primed quantities. Defining \be
W[q(\cd),p(\cd)|q'(\cd),p'(\cd)]=e^{i\tilde{S}[q,p|q',p']}\ee
with: \be
W_0[q(\cd),p(\cd)|q'(\cd),p'(\cd)]=e^{i\tilde{S}_0[q,p|q',p']}\equiv
e^{iS[q(t),p(t)]-iS[q'(t),p'(t)]}, \ee the result, up to second
order in $\l$, is given by:\bea
\lefteqn{W[q(\cd),p(\cd)|q'(\cd),p'(\cd)]=\exp\Big{\{}-\frac{i\l}{4!}\int
dt\,\Big{(}q(t)^4-q'(t)^4\Big{)}}\nn\\&&
+\fr\left(-\frac{i\l}{4!}\right)^2\int dtdt'\,\Big{(}
192iq(t)\De^+(t-t')^3q'(t')+144q(t)^2\De^+(t-t')^2q'(t')^2\nn\\&&-32
iq(t)^3\De^+(t-t')q'(t')^3\Big{)}+0(\l^3)\Big{\}}W_0[q(\cd),p(\cd)|q'
(\cd),p'(\cd)].\eea

We can see that the relationship between the CTP generating
functional and the decoherence functional provides a powerful tool
for the implementation of perturbative techniques in the histories
formalism. We have derived an expression to second order for the
phase space distribution. This is important in two ways. First,
because we have demonstrated an effective, generic method for
dealing with interacting histories theories that can be extended
to any system of interest eg. $\phi^4$ theory, QED. One just
starts from the construction of the CTP generating functional.
Second, because the result is in terms of a distribution on phase
space, we know how to implement coarse-grainings pertaining to a
wide class of interesting systems. This could be of immense use,
for example, in the discussion of how hydrodynamic variables and
their equations of motion emerge from the underlying quantum
theory.

\section{Conclusion}\label{sec:conclusion}

The aims of this paper have been twofold: to complete the phase
space picture of quantum mechanical histories and to demonstrate a
method for the implementation of perturbation theory in the
histories formalism.

We have studied the simpler, quantum mechanical systems of the
simple harmonic oscillator (SHO) the anharmonic oscillator (AHO)
with a quartic potential. Apart from the insights gained in the
present work, these also act as useful pointers to the form that
their field-theoretic analogues may take - the AHO is the natural
quantum mechanical analogue of the self-interacting scalar field
with a $\phi^4$ potential term. In fact, as we shall show in
subsequent work, the results contained herein generalise in a
straightforward manner - modulo renormalization - to field theory.

The analysis of the different phase space representations
complements the work in \cite{Ana00a} and \cite{Ana02}. Although
there exists a continuous infinity of maps between phase space
c-number functions and Hilbert space operators, the most commonly
encountered ones are the Wigner representation and a mapping based
on coherent states. We have explicitly compared these in the
histories formalism, for a general initial state, at both the
discrete-time and continuous-time levels for the case of the SHO.
Ultimately it is the former that are the most important, as all
expressions, (eg., functional integrals involving the
continuous-time expressions) must be understood as a suitable limit
of the mathematically well-defined discrete-time expressions. One
should emphasise that proper implementation of coarse-graining
operations (which is an essential part of the consistent histories
programme) relies on the proper discrete-time expression for the
histories. We concluded that the Q representation was the most
appropriate for phase space histories, chiefly because it allows for
a `Markov' property for the distribution.

Finally, we demonstrated the construction of both the configuration
space and phase space decoherence functional of the AHO. We made use
of the fact that the decoherence functional is related to the
closed-time-path (CTP) generating functional via a Fourier
transform, and that the latter has a well-defined perturbative
expansion.

\bigskip

\noindent\large\textbf{ACKNOWLEDGEMENTS} \normalsize

I would like to thank Ntina Savvidou and Charis Anastopoulos for
their help, advice and support during this work. Thanks are due also
to Chris Isham, Jonathan Halliwell, Howard Carmichael, and Tom ter
Elst for helpful discussions, and to the Physics Department of the
University of Auckland who have put me up as a visitor during the
completion of this work.

\bigskip
\bigskip

\appendix
\Large\textbf{Appendix}\normalsize

\subsection*{The continuum limit of the Wigner representation of the
decoherence functional}

We are calculating the continuum limit of Eq. ({\ref{eq:mtqpdw}). We
take `$n,m$' even. The first exponent is rewritten as\\
$-2\sum_{i=1}^n|\a_i|^2+4\sum_{i=1}^{n-1}\sum_{j=1}^i(-1)^{j+1}e^{-i\o
(t_{i+1-j}-t_{i+1})}\a_{i+1}\a^*_{i+1-j}$

and, following the
derivation in \cite{SA06}, this can be shown to equal\\
$2\sum_{i=1,3}^{n-1}[\a^*_ie^{-i\o t_i}(\a_{i+1}e^{i\o
t_{i+1}}-\a_ie^{i\o t_i})-\a_{i+1}e^{i\o t_{i+1}}(\a^*_{i+1}e^{-i\o
t_{i+1}}-\a^*_ie^{-i\o t_i})]-
\\4\sum_{i=1,3}^{n-1}(\a^*_{i+1}e^{-i\o t_{i+1}}-\a^*_ie^{-i\o
t_i})\sum_{j=i+1,i+3}^{n-2}(\a_{j+2}e^{i\o t_{j+2}}-\a_{j+1}e^{i\o
t_{j+1}})$.

If we now take $n$ large, and thus write
$t_{i+1}-t_i=\d t$, we can rewrite this as\\
$2\sum_{i=1,3}^{n-1}[\a^*_i(\a_{i+1}e^{i\o \d
t}-\a_i)-\a_{i+1}(\a^*_{i+1}-\a^*_ie^{i\o \d t})]-
4\sum_{i=1,3}^{n-1}e^{-i\o t_i}(\a^*_{i+1}e^{-i\o \d
t}-\a^*_i)\sum_{j=i+1,i+3}^{n-2}e^{i\o t_{j+1}}(\a_{j+2}e^{i\o \d
t}-\a_{j+1})$.

Following the same prescription for taking the continuum limit as in
Section \ref{sec:cthpsrQ}, and noting that, since the time steps are
in two's, we will pick up a factor of `1/2' for each
$\sum\rightarrow\int$, we arrive at\\
$\int_{t_1}^{t_n}dt~(\a^*(t)\dot{\a}(t)-\a(t)\dot{\a}^*(t)+2i\o\a^*(t)
\a(t))-
\int_{t_1}^{t_n}dt\int_{t}^{t_n}dt'~e^{-i\o(t-t')}(\dot{\a}^*(t)-i\o
\a^*(t))(\dot{\a}(t')+i\o\a(t'))$.

In the second exponent of Eq. ({\ref{eq:mtqpdw}), ie. the boundary
terms, we note that the first term vanishes as we have taken `$n$'
to be even. We can rewrite the second term (and likewise the third
term)\\ $2\b\sum_{i=1}^n(-1)^{n-i}\a^*_ie^{-i\o t_i} =
2\b\sum_{i=1,3}^{n-1}(\a^*_{i+1}e^{-i\o t_{i+1}}-\a_i^*e^{-i\o
t_i})$.

Thus the second exponent, in the continuum limit is given by \\
$\b\int_{t_1}^{t_n}dt~e^{-i\o
t}(\dot{\a}^*(t)-i\o\a^*(t))-\b^*\int_{t_1}^{t_n}dt~e^{i\o
t}(\dot{\a}(t)+i\o\a(t))$.

We thus arrive at
\\
$\lefteqn{W^{Wigner}[\a(\cd),\a^*(\cd)]=\exp\Big{\{}\int_{t_1}^{t_n}
dt~\Big{(}\a^*(t)\dot{\a}(t)-\a(t)\dot{\a}^*(t)+2i\o\a^*(t)\a(t)
\Big{)}-}$\\
$\int_{t_1}^{t_n}dt\int_{t}^{t_n}dt'~e^{-i\o(t-t')}(\dot{\a}^*(t)-i
\o\a^*(t))(\dot{\a}(t)+i\o\a(t))\Big{\}}\times
\exp\Big{\{}\b\int_{t_1}^{t_n}dt~e^{-i\o
t}(\dot{\a}^*(t)-i\o\a^*(t))-\b^*\int_{t_1}^{t_n}dt~e^{i\o
t}(\dot{\a}(t')+i\o\a(t'))\Big{\}}$.

This expression is significantly refined using integration by
parts and collecting boundary terms to give \\
$W^{Wigner}[\a(\cd),\a^*(\cd)]=\exp\Big{\{}\fr|\a_1|^2-\fr|\a_n|^2-\a^*_1(t_1)\a_n(t_n)-
\int_{t_1}^{t_n}dt~\big{(}\a(t)\dot{\a}^*(t)-i\o\a^*(t)\a(t)\big{)}\Big{\}}$.


\begin{thebibliography}{99}

\bibitem{HIS1} R.~B.~Griffiths. Consistent histories and the
Interpretation of Quantum Mechanics. \emph{J.~Stat.~Phys.}
36:219-272, 1984; R.~Omn\`{e}s. Logical Reformulation of Quantum
Mechanics 1 Foundations. \emph{J.~Stat.~Phys.} 53:893-932, 1988;
M.~Gell-Mann and J.~B.~Hartle. Quantum Mechanics in the light of
Quantum Cosmology. In Complexity, Entropy and the Physics of
Information, ed. W.~Zurek. Addison-Wesley, Reading, 1990.

\bibitem{HPOa} C.~J.~Isham. Quantum Logic and the Histories
Approach to Quantum Theory. \emph{J.~Math.~Phys.} 35:2157, 1994.
gr-qc/9308006; C.~J.~Isham and N.~Linden. The Classification of
Decoherence Functionals: An Analogue of Gleason's Theorem.
\emph{J.~Math.~Phys.} 35:6360-6370, 1994. gr-qc/9406015; C.~J.~Isham
and N.~Linden. Continuous Histories and the History Group in
Generalised Quantum Theory. \emph{J.~Math.~Phys.} 36:5392-5408,
1995. gr-qc/9503063; C.~J.~Isham, N.~Linden, K.~Savvidou and
S.~Schreckenberg. Continuous time and consistent histories.
\emph{J.~Math.~Phys.} 39 (1998) 1818-1834; K.~Savvidou and
C.~J.~Isham. Quantising the Foliation in History Quantum Field
Theory. \emph{J.~Math.~Phys} 43 (2003). quant-ph/0110161.


\bibitem{Sav98} K.~Savvidou. The Action Operator for Continuous
Time Histories. \emph{J.~Math.~Phys.} 40:5657, 1999. gr-qc/9811078.

\bibitem{Sav01a} K.~Savvidou. Poincar\'{e} Invariance for
Continuous-time Histories. \emph{J.~Math.~Phys} 43:3053, 2002.
gr-qc/0104053.

\bibitem{Sav99a} K.~Savvidou. Continuous Time and Consistent
Histories. PhD Thesis, Imperial College, 1999. gr-qc/9912076.

\bibitem{Har93} J.~B.~Hartle. Spacetime Quantum Mechanics and the
Quantum Mechanics of Spacetime. Proceedings on the 1992 Les Houches
School, Gravitation and Quantisation, 1993. gr-qc/9304006.

\bibitem{SavGR03} K.~Savvidou. General Relativity
Histories Theory 1: The spacetime character of the canonical
description. gr-qc/0306034; K.~Savvidou. General Relativity
Histories Theory 2: Invariance groups. gr-qc/0306036.

\bibitem{Bur03} A.~Burch. Histories Electromagnetism.
\emph{J.~Math.~Phys.} 45:2153-2170, 2004. gr-qc/0311092.



\bibitem{Ana00a} C.~Anastopoulos. Continuous-time
Histories: Observables, Probabilities, Phase space Structure and
the Classical Limit. \emph{J.~Math.~Phys.} 42:3225, 2001.
quant-ph/0008052.





\bibitem{Ana02} C.~Anastopoulos. Quantum Processes on Phase Space.
\emph{Annals Phys.} 303:275-320, 2003. quant-ph/0205132.

\bibitem{Sch61} J.~S.~Schwinger. Brownian Motion of a Quantum
Oscillator. \emph{J.~Math.~Phys.} 2:407, 1961.


\bibitem{AW70a} G.~S.~Agarwal and E.~Wolf. Calculus for Functions of
Noncommuting Operators and General Phase-Space Methods in Quantum
Mechanics. I. Mapping Theorems and ordering of Functions of
Noncommuting Operators. \emph{Phys.~Rev.~D.} 2:2161, 1970.


\bibitem{Sri77} M.~D.~Srinivas. Quantum Mechanics as a Generalised
Stochastic Process on Phase Space. \emph{Phys.~Rev.~D.} 15:2837,
1977.

\bibitem{SA06}L.~C.~dos Santos and M.~A.~M.~. de Aguiar. Coherent
State Path Integrals in the Weyl Representation.
\emph{J.~Phys.~A:~Math.~Gen.} 39:13465-13482, 2006.
quant-ph/0607136.

\bibitem{AS06} C.~Anastopoulos and K.~Savvidou. Quantum
Probabilities for Time-extended Measurements. quant-ph/0609021.

\bibitem{CH86} E.~Calzetta and B.~L.~Hu. Closed Time Path
Functional Formalism in Curved Spacetime: Application to
Cosmological Backreaction Problems. \emph{Phys.~Rev.~D.} 35:495,
1987.

\bibitem{CH94} E.~Calzetta and B.~L.~Hu. Noise and Fluctuations in
Semi-classical Gravity. \emph{Phys.~Rev.~D.} 49:6636, 1994.
gr-qc/9312036.

\end{thebibliography}
\end{document}